\documentclass[12pt,letterpaper,hyper]{JHEP3}

\usepackage[latin1]{inputenc}
\usepackage{amsmath,epsfig}
\usepackage{amssymb,amsfonts}
\usepackage{graphicx}
\title{Counting {\it all} dyons in $\CN =4$ string theory}

\preprint{}
\author{
Atish Dabholkar$^{~1, ~2}$, Jo\~ao Gomes$^{~2}$ and Sameer Murthy$^{~2}$\\
\it $^1$Department of Theoretical Physics\\
\it Tata Institute of Fundamental Research\\
\it Homi Bhabha Rd, Mumbai 400 005, India\\

\it $^2${Laboratoire de Physique Th\'eorique et Hautes Energies (LPTHE)\\
\it{Universit\'e Pierre et Marie Curie-Paris 6; CNRS UMR 7589}\\
\it{Tour 24-25, 5$^{\grave{e}me}$ \'etage, Boite 126, 4 Place Jussieu} \\
\it {75252 Paris Cedex 05, France}}\\

}

\abstract{For dyons in heterotic string theory compactified on a six-torus, with  electric charge vector $Q$ and magnetic charge vector $P$, the positive integer  $I \equiv \gcd(Q\wedge P)$  is an invariant of the U-duality group. We propose the microscopic theory for computing the spectrum of \textit{all} dyons for all values of $I$, generalizing earlier results that exist only for the simplest case of $I=1$. Our derivation uses a combination of arguments from duality, 4d-5d lift, and a careful analysis of fermionic zero modes. The resulting degeneracy agrees with the black hole degeneracy for large charges and with the degeneracy of field-theory dyons for small charges. It naturally satisfies several physical requirements including integrality and duality invariance. As a byproduct, we also derive the microscopic $(0,4)$ superconformal field theory relevant for computing the spectrum of five-dimensional Strominger-Vafa black holes in ALE backgrounds and count the resulting degeneracies. }

\keywords{black holes, superstrings, dyons}

\newenvironment{myenumerate}{
\begin{enumerate}
   \setlength{\itemsep}{1pt}
   \setlength{\parskip}{0pt}
   \setlength{\parsep}{0pt}}{\end{enumerate}}

\renewcommand{\Im}{\mbox{Im}}
\renewcommand{\Re}{\mbox{Re}}

\newcommand{\IR}{\mathbb{R}}
\newcommand{\IZ}{\mathbb{Z}}

\newcommand{\Tr}{\mbox{Tr}}

\newcommand{\CC}{\cal{C}}

\def\r{\rho}

\def\s{\sigma}

\def\t{\tau}

\def\h{\eta}

\def\G{\Gamma}

\def\CH{{\cal H}}

\def\CZ{{\cal Z}}
\def\CN{{\cal N}}

\def\half{{\frac12}}

\def\IC{{\mathbb C}}

\def\CN{{\cal N}}

\def\CZ{{\cal Z}}

\def\bea{\begin{eqnarray}}
\def\eea{\end{eqnarray}}
\def\be{\begin{equation}}
\def\ee{\end{equation}}
\def\ba{\begin{align}}
\def\ea{\end{align}}
\def\bse{\begin{subequations}}
\def\ese{\end{subequations}}
\def\1F1{{}_1\!F_1}
\def\2F0{{}_2\!F_0}

\def\G{\Gamma}

\def\h3{$\textrm{H}_3^+$}
\def\IC{{\mathbb C}}
\def\IR{{\mathbb R}}
\def\IZ{{\mathbb Z}}

\newcommand{\beq}{\begin{equation}}
\newcommand{\eeq}{\end{equation}}
\newcommand{\ber}{\begin{eqnarray}}
\newcommand{\eer}{\end{eqnarray}}

\def\be{\begin{eqnarray}}
\def\ee{\end{eqnarray}}

\def\mod{{\rm mod}}

\def\CN{{\cal N}}

\def\CZ{{\cal Z}}
\def\CE{{\cal E}}

\def\CH{{\cal H}}
\def\CC{{\cal C}}

\def\CC{{\cal C}}

\def\CE{{\cal E }}

\def\CZ{{\cal Z }}

\def\Tr{{\rm Tr}}

\def\G{\Gamma}

\font\manual=manfnt
\def\dbend{\lower3.5pt\hbox{\manual\char127}}

\def\bar{\overline}

\def\CN{{\cal N}}
\def\CH{{\cal H}}
\def\rt2{\sqrt{2}}
\def\irt2{{1\over\sqrt{2}}}

\def\t{\tilde}

\def\s{\sigma}

\def\mod{{\rm mod}}

\font\cmss=cmss10
\font\cmsss=cmss10 at 7pt
\def\IL{\relax{\rm I\kern-.18em L}}
\def\IH{\relax{\rm I\kern-.18em H}}
\def\rlx{\relax\leavevmode}
\def\ZZ{\rlx\leavevmode\ifmmode\mathchoice{\hbox{\cmss Z\kern-.4em Z}}
 {\hbox{\cmss Z\kern-.4em Z}}{\lower.9pt\hbox{\cmsss Z\kern-.36em Z}}
 {\lower1.2pt\hbox{\cmsss Z\kern-.36em Z}}\else{\cmss Z\kern-.4em
 Z}\fi}
\def\Tr{{\rm Tr}}

\def\G{\Gamma}

\def\rt2{\sqrt{2}}
\def\irt2{{1\over\sqrt{2}}}

\def\t{\tilde}
\def\T{\widetilde}

\def\s{\sigma}

\begin{document}

%\maketitle \setcounter{tocdepth}{2}
%\tableofcontents

\section{Introduction}

In this note we propose the microscopic theory for counting all possible dyons in  heterotic string theory compactified on a six-dimensional torus $T^6$. The resulting four-dimensional theory has  $\mathcal{N}=4$ supersymmetry with a U-duality group
\begin{equation}\label{dulitygroup}
   G(\mathbb{Z}) \equiv O(22, 6; \mathbb{Z}) \times SL(2, \mathbb{Z}).
\end{equation}
A dyonic state in the theory is specified by a  charge vector
\begin{equation}\label{chargevector}
    \Gamma^i_\alpha \equiv \left[
                             \begin{array}{c}
                               Q^i \\
                               P^i \\
                             \end{array}
                           \right]
\end{equation}
where the index $i = 1, \ldots, 28$ transforms in the  vector representation of the T-duality group $O(22, 6; \mathbb{Z})$ and $\alpha = 1, 2$ transforms in the fundamental representation of the S-duality group $SL(2, \mathbb{Z})$. The components $Q^i$ and $P^i$ can be regarded as the electric and magnetic charge vectors of the state respectively. A quarter-BPS dyon is characterized by the following relation between the mass $M$ and  the two central charges $Z_1(\Gamma, \phi)$ and $Z_2(\Gamma, \phi)$ of the $\CN=4$ superalgebra:
\begin{equation}\label{mass}
    M = |Z_1(\Gamma, \phi)| > |Z_2(\Gamma, \phi)|.
\end{equation}
The mass thus depends both on the charge $\Gamma$ of the state and the asymptotic values of the moduli which we have denoted generically by $\phi$. Such a state preserves four of the sixteen supersymmetries and belongs to a $64$-dimensional supermultiplet. We will be interested in knowing the degeneracy $\Omega(\Gamma, \phi)$ of all such dyons in the theory.

One physical requirement on the dyon spectrum is that it should be duality invariant. More precisely, this means that if $(\Gamma, \phi)$ transform to $(\Gamma', \phi')$ under a duality transformation, then we must have $\Omega(\Gamma', \phi') = \Omega(\Gamma, \phi)$. It is therefore useful to know the duality invariants that one can form from the charge vector $\Gamma$.
There is a unique quartic invariant of the duality group
 \begin{equation}\label{Delta}
    \Delta = Q^2 P^2 -(Q\cdot P)^2,
 \end{equation}
which is an invariant not only of $G(\mathbb{Z})$ but also of $G(\mathbb{R})$, the continuous form of the duality group over real numbers. Since $G(\mathbb{R})$ is a symmetry of the low energy supergravity action, macroscopic physical quantities which follow from the two-derivative supergravity action, such as the leading Bekenstein-Hawking entropy of the corresponding black holes, are expected to be functions of this duality invariant $\Delta$.

In the full quantum theory, because of the Dirac quantization condition, only the discrete duality group $G(\mathbb{Z})$ is  a symmetry that preserves the integrality of charges. Since this is a smaller symmetry than the  continuous one, there are more subtle additional invariants. For example,  one such important invariant  noted in \cite{Dabholkar:2007vk} in this context is defined by
\begin{equation}\label{invariant}
    I = \gcd (Q \wedge P).
\end{equation}
This positive integer is an invariant of $G(\mathbb{Z})$ but not of $G(\mathbb{R})$, because $\textrm{g.c.d.}$ is a concept defined only for integers and not for reals.
We refer to such invariants as `discrete invariants' to underscore the fact that they are invariants of the discrete group but not of the continuous one.

As we will discuss later in  more detail, $I$ is essentially the unique discrete invariant that is relevant to this problem. The duality orbits of dyons can then be subdivided into an infinite number of families labeled by the positive integer $I$. Much of the earlier work starting with the work of Dijkgraaf, Verlinde, Verlinde
concerns the spectrum of dyons in only the first of these families with $I=1$ \cite{Dijkgraaf:1996it, Gaiotto:2005gf, {Shih:2005uc}, {Shih:2005he}, {Dabholkar:2006xa}, David:2006yn,{Gaiotto:2005hc}}. Generalizations to other $\mathcal{N} =4$ orbifolds are also restricted to the case $I=1$ \cite{{Jatkar:2005bh}, {David:2006ji}, {Dabholkar:2006xa}, {Dabholkar:2007vk}, {David:2006yn}, David:2006ru, {Dabholkar:2006bj}}. States which have the same value $\Delta$ but different value of $I$ will in general have different degeneracies because they belong to different duality orbits. For black holes, this will show up  at the exponentially subleading order. But as we will see, these contributions are important for comparison with field theory. Our objective will be to compute the degeneracy $\Omega(\Gamma, \phi)$ for all values of  $I$.

In section $\S{\ref{Duality}}$, we begin with some generalities about duality and its implications for the dyon spectrum. In $\S{\ref{Discrete}}$ we review the invariants that characterize the  dyons, in particular the discrete ones, and discuss the expected transformation properties of the dyon spectrum.  We then  discuss the basic properties of the dyon partition function in  terms of the Siegel modular forms in $\S{\ref{General}}$.

In section $\S{\ref{Microscopic}}$, we turn to the microscopic derivation. In $\S{\ref{Rep}}$, we consider representative members of dyons in each duality orbit both in heterotic and Type-II frame. In particular,  we choose a system of  D-branes and multiple Kaluza-Klein monopoles. In $\S{\ref{Multiple}}$, we  review the relevant aspects of the geometry of Kaluza-Klein monopoles and discuss the 4d-5d lift in this context. In $\S{\ref{Five}}$ we address the simpler but closely related problem of counting the degeneracies of five dimensional three charge black holes in ALE space. In the appendix $\S{\ref{Multiple}}$, we analyze the problem of bound states of KK-P system by mapping it to the F1-P system paying special attention to the fermionic zero modes.
Using the insights from the five-dimensional case and the KK-P bound states, we discuss  the degeneracies of four-dimensional dyons in $\S{\ref{Four}}$.  We propose the microscopic $(1+1)$-dimensional superconformal field theory relevant for this counting. Using this SCFT, we derive the partition function for the dyons.

In section $\S{\ref{Consistency}}$, we discuss the modular properties of the partition function and their physical consequences. We show in $\S{\ref{Duality2}}$ that the  partition function obtained from the microscopic theory is  given in terms of  Siegel modular forms that are invariant under $\Gamma^0(I)$ subgroup of $Sp(2, \mathbb{Z})$. We show how to extract the degeneracies from this partition function  that exhibit the duality invariance under the full $SL(2, \mathbb{Z})$ S-duality group along with an expected dependence on the moduli. In $\S{\ref{Comparison}}$, we consider the field theory limit of these string theory dyons and show that the spectrum is in agreement with independent field theoretic computations. Our microscopic partition function is in agreement  with recent macroscopic proposal in \cite{Banerjee:2008pu, {Banerjee:2008pv}} which was motivated from the analysis of various physical constraints on the spectrum of dyons.

We conclude in $\S{\ref{Conclusions}}$ with discussion and comments.

\section{Duality and the dyon spectrum \label{Duality}}

In this section we consider the implications of duality  for formulating the dyon partition functions in different duality orbits.

\subsection{Duality invariants for $\mathcal{N}=4$ dyons\label{Discrete}}

Both $Q^i$ and $P^i$ are lattice vectors in the self-dual, even, Lorentzian Narain lattice $\Lambda^{22, 6}$ on which  $O(22, 6; \mathbb{Z})$ has a natural action. Using $O(22, 6; \mathbb{R})$ invariant Lorentzian metric
\begin{equation}\label{lorentzian}
    L_{ij} = \left(
               \begin{array}{ccc}
                 - \textbf{1}_{16 \times 16} & \textbf{0} & \textbf{0}\\
                 \textbf{0} & \textbf{0} & \textbf{1}_{6 \times 6}  \\
                 \textbf{0} &   \textbf{1}_{6 \times 6} & \textbf{0}  \\
               \end{array}
             \right),
\end{equation}
one can define three T-duality invariants  $Q^2$, $P^2$, and $Q \cdot P$. Using the antisymmetric tensor $\epsilon_{\alpha\beta}$ of $SL(2, \mathbb{R})$, we can write the U-duality invariant $\Delta$ as \begin{equation}\label{Delta2}
    \Delta = L_{ij} L_{kl} \epsilon_{\alpha\beta} \epsilon_{\delta\gamma} \Gamma^i_\alpha \Gamma^j_\beta \Gamma^k_\delta \Gamma^l_\gamma
\end{equation}
which is manifestly an invariant also of the continuous group $G(\mathbb{R})$.

Turning to the discrete invariant, there are a number of ways to see that the positive integer  $I$ defined in (\ref{invariant}) is an invariant of $G(\mathbb{Z})$. Geometrically, given two vectors $Q$ and $P$  in the Narain lattice, $(Q\wedge P)$ defines the area tensor of  the parallelogram  spanned by them. The invariant $I-1$ then counts the number of lattice points inside this parallelogram \cite{Dabholkar:2007vk}. If $I=1$, then the parallelogram is a primitive cell of the lattice, otherwise it is non-primitive.
There is also a group-theoretic way to see that $I$ is an invariant \cite{Banerjee:2007sr}. We can write
\begin{equation}\label{I2}
    I = \gcd (\epsilon_{\alpha\beta} \Gamma^i_\alpha \Gamma^j_\beta).
\end{equation}
This positive integer is manifestly $SL(2, \mathbb{Z})$ invariant. Furthermore, the area tensor $(\epsilon_{\alpha\beta} \Gamma^i_\alpha \Gamma^j_\beta)$ transforms linearly in the antisymmetric tensor representation of $G(\mathbb{Z})$ which is represented by  matrices with integer entries. Linear multiplication by an invertible integral matrix of $O(22, 6; \mathbb{Z})$ cannot change the greatest common divisor  of all  components of the area tensor, and hence $I$ is an invariant.

Subsequent analysis \cite{{Banerjee:2007sr}, Banerjee:2008ri} revealed that $I$ is essentially the only discrete invariant that we need to worry about in this context. To see this, let us first analyze the discrete invariants of the T-duality group $O(22, 6; \mathbb{Z})$. We assume that the total charge vector $\Gamma = (Q, P)$ is  primitive so that it cannot be written as a multiple of any other  lattice vector $(Q_0, P_0)$. Otherwise, the dyon can split into a number of single particle states. Now, even  if we restrict $\Gamma$ to be  primitive, $Q$ and $P$ can individually be nonprimitive. This means for example that we can write $Q =r_1 Q_0$ and $P = r_2 P_0$ for some lattice vectors $Q_0$ and $P_0$, where $r_1$ and $r_2$ are integers without a common factor. Now, since we are interested in quarter-BPS dyons, $Q$ and $P$ must be nonparallel. Hence, the two vectors $Q$ and $P$ generate a two-dimensional sublattice $\Lambda_0$ of the Narain lattice. It is then possible to choose a basis for this sublattice $(e_1, e_2)$ \cite{Banerjee:2007sr} so that
\begin{eqnarray} \label{tdualityinv}
  &&Q = r_1 e_1, \quad  P = r_2 (u_1e_1 + r_3 e_2), \quad r_1, r_2, r_3, u_1 \in \mathbb{Z}^+\\
  &&\gcd(r_1, r_2) = \gcd(u_1, r_3)  = 1 \quad 1\leq u_1 \leq r_3.
\end{eqnarray}
The integers $r_1$, $r_2$, $r_3$, $u_1$ together with $Q^2$, $P^2$, and $Q \cdot P$ can be shown to be the complete set of T-duality invariants \cite{Banerjee:2007sr}.

Given the T-duality invariants, one can then investigate the consequences of S-duality invariance. Note that the discrete U-duality invariant $I$ is simply the product $ r_1 r_2 r_3$. It can be shown  \cite{Banerjee:2008ri, {Nampuri:2007gv}} that  a general set of discrete invariants $(r_1, r_2, r_3, u_1)$ can be mapped to a representative in this orbit of the form $(I, 1, 1, 1)$ by an $SL(2, \mathbb{Z})$ transformation. Moreover, this choice of the representative is invariant under the congruence subgroup $\Gamma^0(I)$ of the $SL(2, \mathbb{Z})$.
It will therefore suffice to compute the degeneracy in each family with the discrete invariants fixed to $(I, 1, 1, 1)$, which is expected to exhibit the $\Gamma^0(I)$ symmetry.

\subsection{Generalities about the dyon partition function \label{General}}

We will see that for each orbit labeled by the discrete invariant $I$, the degeneracies of dyons are summarized in terms of a dyon partition function $\mathcal{Z}_{\rm I}(\rho, \sigma, v)$ of  three complex variables
$(\rho, \sigma, v)$, which  can be thought of as the three (complexified) chemical potentials for the three T-duality invariant integers $(Q^2/2, P^2/2, Q\cdot P)$ respectively.

One of the unexpected aspects of the dyon partition function is that it can be expressed in terms of certain  Siegel  modular forms of $Sp(2, \mathbb{Z})$ and its congruence subgroups. This large modular symmetry allows us to demonstrate S-duality invariance of the spectrum,  but the full $Sp(2, \mathbb{Z})$ symmetry itself is too large to be accommodated inside the physical duality group $G(\mathbb{Z})$ in any obvious way.
It has become clear recently that this symmetry has other important physical consequences. For example, it determines elegantly the moduli dependence of the spectrum  and the structure of walls of marginal stability \cite{{Dabholkar:2007vk}, {Sen:2007vb}, Cheng:2007ch}.
The physical origin of this symmetry for the special case $I=1$ can be explained from the representation of the dyons in M-theory using M5-branes \cite{Gaiotto:2005hc, Dabholkar:2006bj} but the situation for higher values of $I$ remains mysterious.

To establish the notation, let us recall a few facts about Siegel modular forms. The three chemical potentials can be packaged together as a symmetric  $(2\times 2)$ matrix $\tau$  with complex entries
\begin{equation}\label{period}
   \tau = \left(
              \begin{array}{cc}
                \rho & v \\
                v & \sigma \\
              \end{array}
            \right)
\end{equation}
satisfying
\begin{equation}\label{cond1}
   (\textrm{Im} \rho) > 0, \quad (\textrm{Im} \sigma) > 0, \quad
   (\textrm{Im} \rho)(\textrm{Im} \sigma) > (\textrm{Im} v)^2
\end{equation}
which parametrizes  the `Siegel upper half-plane' in the space of
$(\rho, v, \sigma)$.  We write an element $g$ of $Sp(2, \mathbb{Z})$ as a
$(4\times 4)$ matrix in the block form as
\begin{equation}\label{sp}
   \left(
  \begin{array}{cc}
    A & B \\
    C & D \\
  \end{array}
\right),
\end{equation}
where $A, B, C, D$ are all $(2\times 2)$ matrices with integer
entries. They  satisfy
\begin{equation}\label{cond}
   AB^T=BA^T, \qquad  CD^T=DC^T, \qquad AD^T-BC^T= \mathbf{1}\, ,
\end{equation}
so that $g J g^t =J$ where $J = \left(
                                  \begin{array}{cc}
                                    0 & -\mathbf{1} \\
                                    \mathbf{1} & 0 \\
                                  \end{array}
                                \right)$
is the symplectic form. The matrix $\tau$  can be thought of as the  period matrix of a genus two Riemann surface on which there is a natural symplectic action of $Sp(2, \mathbb{Z})$. An element $g$ of the form (\ref{sp}) acts as
\begin{equation}\label{trans}
    \tau \to (A \tau + B )(C\tau + D ) ^{-1}.
\end{equation}
With these definitions, genus-two Siegel modular forms can be defined as a generalization of genus-one modular forms that transform under $SL(2, \mathbb{Z}) \sim Sp(1, \mathbb{Z})$. A Siegel modular form $\Phi_k(\tau)$ of weight $k$ is holomorphic in the Siegel upper half-plane and transforms under the transformation (\ref{trans}) as
\begin{equation}\label{phi}
    \Phi_k [(A \tau + B )(C\tau + D ) ^{-1}] =  \{\det{(C\tau + D )}\}^k
    \Phi_k (\tau).
\end{equation}

It turns out that the partition functions can be expressed compactly  in terms of Siegel modular forms. For example, in the simplest case $I=1$, the partition function $\mathcal{Z}_1(\tau)$ is given by
\begin{equation}\label{igusa}
    \mathcal{Z}_1(\tau) = \frac{1}{\Phi_{10}},
\end{equation}
where $\Phi_{10}$ is the well-known Igusa cusp form, which is the unique weight $10$ form of $Sp(2, \mathbb{Z})$. The partition function therefore transforms as in (\ref{trans}) with a negative weight $-10$.  There is a natural embedding of the physical S-duality group $SL(2, \mathbb{Z})$ into $Sp(2, \mathbb{Z})$,  which we will discuss in detail in $\S{\ref{Duality2}}$. Under this embedding the $SL(2, \mathbb{Z})$ are represented by certain $Sp(2, \mathbb{Z})$ matrices with $C=0$.  Combining this fact with the modular property (\ref{trans})
one can then conclude that the partition function is $SL(2, \mathbb{Z})$ \emph{invariant}.

Note that the partition function is not necessarily holomorphic and can have poles in the Siegel upper half-plane. Indeed, these poles in the partition function imply  `phase transitions' which in this supersymmetric context correspond to crossing walls of marginal stability. On the other hand, as one might expect on physical grounds,  the partition function  has no zeros as a function of the chemical potentials, which follows from the  fact that $\Phi_{10}$ has no poles in the Siegel upper half-plane.

As we will see, the structure generalizes naturally for $I >1$. In all cases, the partition function $\CZ_I$ is invariant under a congruence subgroup $\Gamma^0(I) \subset SL(2, \mathbb{Z})  \subset Sp(2, \mathbb{Z})$. In terms of this partition function, the degeneracy is then given by \footnote{The subscript $I$ is not strictly necessary for $\Omega$ once the charges are specified, but we have added it nevertheless to  emphasize the orbit to which the dyons belong. }
\begin{equation}\label{inverse3}
   \Omega_{I}(\Gamma, \phi) = (-1)^{P\cdot Q +1}\int_{\CC}
   d^3 \tau \, e^{-i\pi \Gamma^{t} \cdot
     \tau \cdot \Gamma}\,
   {\CZ_{I}(\tau)}
\end{equation}
where the integral is over the contour
\begin{eqnarray}\label{contour}
  &&0 < {\Re(\rho)} \leq 1, \quad 0 < \Re(\sigma) \leq 1, \quad 0 < \Re(v ) \leq 1 \\
  && {\Im(\rho)} >>1, \quad \Im(\sigma) >>1, \quad \Im(v )>> 1
\end{eqnarray}
over the three coordinates, where $\Re$ and $\Im$ denote the real and imaginary parts.
This defines the integration curve $\CC$ as a 3-torus in the Siegel
upper half-plane. The choice of the contour is defined by the precise values of the imaginary parts and is determined by the region of the moduli space to which $\phi$ belongs. This moduli dependence and other physical properties of the partition function will be discussed later
in $\S{\ref{Consistency}}$. Note that the degeneracy defined above is  really an index computing the difference between 64-dimensional supermultiplets built on either bosonic or fermionic ground states.

\section{Microscopic derivation \label{Microscopic}}

We now turn to the microscopic derivation of the degeneracies. Following the experience  for the $I=1$ case, we will make use of the 4d-5d lift \cite{{Gaiotto:2005gf}, Shih:2005uc} to relate this four-dimensional computation to a five dimensional computation of  D1D5P black holes. One novelty for $I >1$ is that the counting involves multiple Kaluza-Klein monopoles and the superconformal field theory (SCFT) on the effective string describing this counting is more complicated. We propose an effective SCFT using an analysis of D1-D5 brane in ALE space, of multi-particle states of multiply wound strings and their fermion zero modes, and constraints  from duality invariance. The microscopic partition function is  then given in terms of a modified elliptic genus of this SCFT.

\subsection{Representative configurations of dyons \label{Rep}}

To prepare for the microscopic derivation, we would like to choose simple representatives in each duality orbit. We first label the charges in the heterotic frame where the $SL(2, \mathbb{Z})$ duality group corresponds to the electric-magnetic S-duality. We then go to a Type-IIB frame to map the configuration to  a system of D-branes.

Consider heterotic string theory on $\mathbb{R}^{1, 3} \times {\tilde S}^1  \times S^1 \times   T^4 $. The noncompact Minkowski spacetime $\mathbb{R}^{1, 3}$ has coordinates $X^\mu$ with $ \mu = 0, 1, 2, 3$. The circles  ${\tilde S}^1$  and $S^1$ have coordinates $X^4$  and $X^5$ respectively. The torus $T^4$ has coordinates $X^m$ with  $m = 6, 7, 8, 9$.

Let  $n$, $w$, $K$, and $W$ be the momentum, winding, KK-monopole, and NS5-brane charges respectively associated with the circle $S^1$.
Similarly $\tilde n$, $\tilde w$, $\tilde K$, and $\tilde W$ are the corresponding charges associated with the circle ${\tilde S}^1$. There are two potentially confusing points about the notation and the physics. First, even though the state with charge $W$ is associated with the $ S^1$ circle and is charged with respect to the 4d gauge field  $B_{\mu 5}$ coming from the reduction of the antisymmetric tensor field with one index along the $S^1$ circle, it corresponds to an NS5-brane wrapping along the ${\tilde S}^1$ circle and the $T^4$. Second, even though $K$ is the charge that is magnetically dual to $n$ is  in terms of Dirac quantization
condition, the state that is S-dual  to $n$ under the $SL(2, \mathbb{Z})$ group is $W$ because of the way the various $\mathbb{Z}_2$ duality symmetries are embedded into the nonabelian duality group $G(\mathbb{Z})$. Similar comments hold for charges with the tilde.  With these notations, we can  consider a dyon which is a bound state of all these objects, so that we have
\begin{equation}\label{hetcharges}
    \Gamma = \left[
               \begin{array}{c}
                 Q \\
                 P \\
               \end{array}
             \right] =
             \left[
               \begin{array}{cccc}
                 {\tilde n},& n ; & {\tilde w},& w \\
                 {\tilde W},& W; & {\tilde K}, & K \\
               \end{array}
             \right]_H,
\end{equation}
where the subscript $H$ denotes that we are labeling the charges in the heterotic frame.

To compute the continuous T-duality invariants of the dyon with these eight charges, we can use  the metric (\ref{lorentzian}) restricted to the $\Lambda^{2, 2}$ Narain lattice associated with the two circles. One then obtains
\begin{equation}\label{invariants}
    Q^2 = 2 (\tilde n \tilde w + nw );\quad  P^2 = 2 (\tilde W \tilde K + WK ); \quad Q\cdot P  = (\tilde n \tilde K + n K + \tilde w \tilde W +  w W ).
\end{equation}
We can thus realize all integer values of these duality invariants using the eight basic charges.
%If we insist that  subject to the constraint $Q^2 \geq -2$ and $P^2 \geq %-2$, which holds from the on-shell Virasoro condition for the electric %charges and by S-duality for the magnetic charges.

For deriving the spectrum it is useful to go to the Type-IIB frame and  map the dyon configuration to a system of D-branes in Taub-NUT geometry of KK-monopoles. This can be achieved in three steps.
\vspace{-2mm}
\begin{myenumerate}
  \item We  first use string-string duality to go to Type-IIA on $\mathbb{R}^{1, 3} \times {\tilde S}^1  \times S^1 \times   K_3 $. Under this duality, the momentum and the KK-monopole charges are not affected. Since a fundamental heterotic string wrapping a circle is an NS5-brane of Type-IIA wrapping the same circle and $K3$, a winding charge $w$ in the heterotic frame is relabeled as the NS5-brane charge $\tilde W$ in the Type-II frame. We denote this  IIA frame with a subscript $A$.
  \item  We then T-dualize along the $\tilde S^1$ circle to go to Type-IIB frame. Under this duality, $\tilde n$ and $\tilde w$ in the IIA frame get relabeled respectively as $\tilde w$ and $\tilde n$ in the IIB frame. Similarly $\tilde K$ and $\tilde W$ in the IIA frame get relabeled as $\tilde W$ and $\tilde K$ respectively. We denote this IIB frame with a subscript $B'$.
  \item  Finally, we use  ten-dimensional S-duality of Type-IIB which maps the NS5-branes and winding strings to D5-branes and D1-branes respectively.  Under this duality, $w$ gets mapped to $Q_1$ which denotes the charge of D1-branes wrapping $S^1$. Since $W$ is wrapping $\tilde  S^1$ and $K3$, it gets mapped to $\tilde Q_5$ which denotes the charge of D5-branes wrapping $\tilde  S^1$ and $K3$. Similarly, $\tilde w$ and $\tilde W$ turn into $\tilde Q_1$ and $Q_5$ respectively\footnote{In our notation $Q_1$ and $Q_5$ denote physical charges. If we denote by $N_1$, $N_5$ the numbers of D1 and D5 branes respectively, then we have $Q_5 =N_5$ but $Q_1 = N_1- N_5$ because of contribution to the D1-brane charge coming from the Euler character of $K3$.}. Momentum and KK charges remain unchanged. We denote this IIB frame with a subscript $B$.
\end{myenumerate}
\vspace{-2mm}
In these three duality frames, the charges above are labeled as
\begin{equation}\label{hetchargesII}
    \Gamma =
             \left[
               \begin{array}{cccc}
                 {\tilde n},& n ; & W, & {\tilde W} \\
                 w,& {\tilde w}; & {\tilde K}, & K\\
               \end{array}
             \right]_A =
             \left[
               \begin{array}{cccc}
                  {\tilde w},& n ; & W, & {\tilde K} \\
                 w,& {\tilde n}; & {\tilde W}, & K\\
               \end{array}
             \right]_{B'} = \left[
               \begin{array}{cccc}
                 {\tilde Q_1},& n ; & {\tilde Q_5}, & {\tilde K} \\
                 Q_1,& {\tilde n}; & Q_5, & K\\
               \end{array}
             \right]_B \, .
\end{equation}
In what follows, we choose the following configuration of charges
\begin{equation}\label{final charges}
    \Gamma = \left[
      \begin{array}{cccc}
                 0,& n ; & 0, & {\tilde K} \\
                 Q_1,& {\tilde n}; & Q_5, & 0\\
       \end{array}
      \right]_B,
\end{equation}
in the IIB frame. The discrete invariant $I$ for this configuration is
\begin{equation}\label{equation}
    I =\gcd (n Q_1, n Q_5, {\tilde K} Q_1, {\tilde K} Q_5, {\tilde K} {\tilde n}).
\end{equation}
It is clear then, that there are two simple ways to obtain an arbitrary value for $I$.
\vspace{-2mm}
\begin{myenumerate}
  \item We can take $\tilde K =I$ and $n = I m$ and choose $Q_1$,  $Q_5$, $m$, $\tilde n$ to be  relatively prime with respect to each other. In the notation of (\ref{tdualityinv}) we then have $r_1 =I$, $r_2 = r_3 =1$.
  \item We can take $Q_1 = I q_1 $,  ${\tilde n} = I {\tilde m}$,  $Q_5 = I q_5$ and  choose $q_1$,  ${\tilde K}$,  $q_5$, $\tilde m$, $n$ to be relatively prime with respect to each other. In the notation of (\ref{tdualityinv}) we then have $r_2 =I$, $r_1 = r_3 =1$.
\end{myenumerate}
\vspace{-2mm}
In the following section, we consider the first configuration above to derive the degeneracies so that the KK-monopole charge is $I$ and the momentum is $n = I m$. This  generalizes the $\tilde K =1$ case which was used for  a 4d-5d lift in \cite{Gaiotto:2005gf,{Shih:2005uc}, {David:2006yn}}. Since this configuration can realize all values of the duality invariant, by deducing the degeneracy for each such representative, we can obtain the fully U-duality invariant spectrum of dyons for all values of $I$.

\subsection{Multiple Kaluza-Klein monopoles and 4d-5d lift\label{Multiple}}

The geometry of $I$ KK-monopoles \cite{{Sorkin:1983ns}, Gross:1983hb} is given by the multi-centered Taub-NUT space $TN_I$ of charge $I$ \cite{Gibbons:1979zt} with metric
\begin{equation}\label{taubnut}
    ds^2_{TN_I} = V^{-1} (dx^4 + \vec{\omega}\cdot  d\vec{x})^2 + V d\vec{x} \cdot d\vec{x},
\end{equation}
where $x^4$ is a compact direction  and $\vec{x} \equiv (x^1, x^2, x^3)$ are coordinates in $\mathbb{R}^3$. The harmonic function $V(\vec{x})$ and the vector potential $\vec\omega$ are defined by
\begin{eqnarray}\label{V}
  V &=& 1 + \sum_{s=1}^I V_s,\qquad \vec{\omega} = \sum_{s=1}^I \vec{\omega_s}\, ; \\
  V_s &=& \frac{4m}{|\vec{x} - \vec{x_s}|},\qquad\quad \vec\nabla \times \vec{\omega_s} =  \vec\nabla  V_s \, .
\end{eqnarray}
At asymptotic infinity  when $|\vec{x}|$ is very large, the geometry  asymptotes to $\mathbb{R}^3 \times \tilde S^1$.
The vectors $\{\vec{x_s}\}$ can be interpreted as the $I$ locations of the KK-monopoles in the transverse $\mathbb{R}^3$ space.  The coordinate $x^4$ must have periodicity $16 \pi m$ in order that the solutions are free from conical singularities at $\vec{x} = \vec{x}_s$. We can therefore identify $8m$ with the radius $\tilde R$ of the circle $\tilde S^1$.

The multi-centered Taub-NUT space supports $I$ linearly independent, self-dual, normalizable harmonic 2-forms $\Sigma_s$ given by \cite{Ruback:1986ag}
\begin{equation}\label{harmonic}
   \Sigma_s = d \sigma_s; \quad \sigma_s =  V^{-1} V_s (dx^4 + \vec{\omega}\cdot  d\vec{x}) - \vec{\omega_s} \cdot d\vec{x},
\end{equation}
normalized as
\begin{equation}\label{norm}
    \int \Sigma_s \wedge \Sigma_t = (16 \pi m)^2 \delta_{st}.
\end{equation}

Let us recall the arguments regarding the 4d-5d lift when $I =1$. In this case, the geometry of a single KK-monopole is given by the Taub-NUT space of unit charge. We can take $\vec{x}_1 =0$ so that the KK-monopole is localized at the origin. This space has a $ U(1)$ translational symmetry along the $x^4$ coordinate and $Spin(3)$ rotational symmetry in $\mathbb{R}^3$. Far away from the origin, $|x| >> \tilde R$, this space asymptotes to $\mathbb{R}^3 \times \tilde S$. Close to the origin, $|x| << \tilde R$ it looks like $\mathbb{R}^4$. The Euclidean space $\mathbb{R}^4$ has $Spin(4) \sim SU(2)_L \times SU(2)_R$ rotational symmetry. Let  $J_{12}$ and $J_{34}$ be the Cartan generators of the $Spin(4)$ corresponding to rotations in the $12$ plane and $34$ plane respectively. Then the Cartan generators  $U(1)_L \times U(1)_R$  of the $ SU(2)_L \times SU(2)_R$ are
\begin{equation}\label{j}
    J^3_L = \half (J_{12} +J_{34}) , \quad J^3_R = \half (J_{12} - J_{34}).
\end{equation}
The $Spin(3)$ symmetry of the full geometry can be identified with $SU(2)_R$ symmetry at the origin and the $U(1)$ translational symmetry along the circle with $U(1)_L$ at the origin.

By taking the radius of the $\tilde S^1$ circle to be large compared to the Planck scale, one can focus on the region near the origin.  Moreover, since the generator of translations along the KK-circle $\tilde S^1$  becomes the generator $2J_L$, we can identify the asymptotic momentum $\tilde n$ with the angular momentum $l$ in $\mathbb{R}^4$ near the core of the Taub-NUT space. This allows one to relate the 4-dimensional dyon to the 5-dimensional D1-D5 system corresponding to the BMPV black hole \cite{Breckenridge:1996is} with momentum $n$ and angular momentum $2J_L =\tilde n = l$. The 5d counting problem of the D1-D5 system is captured by a $(4, 4)$ two-dimensional superconformal field theory along  the worldvolume $\mathbb{R} \times S^1$ with target space $\mathrm{Sym}^{Q_1Q_5 +1}(K3)$ \cite{Vafa:1995bm}. We denote this sigma model SCFT by
\begin{equation}\label{scft}
    \s(\mathrm{Sym}^{Q_1Q_5 +1}(K3) ).
\end{equation}
To count the $4d$ degeneracy, one must also consider contributions from the center of mass (COM) motion of the D1-D5 system in the Taub-NUT space, as well as the bound states of momentum and KK-monopole \cite{{Dabholkar:2006xa},David:2006yn}.

For $I >1$, there are several difficulties in generalizing the above considerations. First, multi-KK monopoles have collective coordinates describing their relative motions which parametrize  a nontrivial moduli space. One needs to understand quantum mechanics on this moduli space to fully understand the spectrum of bound states. This is a potentially complicated problem but using the fact that KK-monopole is dual the heterotic fundamental string, we can predict what kind of bound states are possible. From this analysis and from the analysis  of the  fermionic zero modes, we will argue that for the counting problem of our interest, all KK-monopoles are essentially sitting on top of each other with an appropriately symmetrized wavefunction. Second, the Taub-NUT space admits $I$ nontrivial 2-cycles Poincar\'e dual  to self-dual harmonic 2-forms (\ref{harmonic}). For a pair of KK-monopoles there is a 2-cycle that touches both of them \cite{Sen:1997kz}. The area of this 2-cycle is proportional to the distance between the two monopoles and goes to zero as the monopoles approach each other. A D3-brane wrapping such a cycle will give rise to tensionless strings \cite{Witten:1995zh}.  As a result, while quantizing the moduli space of KK-monopoles, one has to take into account the potential contribution from these tensionless strings which is a difficult problem.  To gain some insight into these difficulties, we will consider two closely related simpler problems.\begin{itemize}
  \item First, in the following subsection $\S{\ref{Five}}$, we consider the situation when all KK-monopoles are sitting at the origin so that $\vec{x}_s =0$ for all $s$  and the radius $\tilde R$ is very large. In this limit, multi-centered Taub-NUT space becomes an asymptotically locally Euclidean  (ALE) space  $\IC^2/\IZ_I$. Moreover, the KK-P bound states move out of the spectrum, and the COM motion of the D1-D5 system is on the ALE space.
  \item Second, in the appendix $\S{\ref{FP}}$, we discuss the multiparticle Hilbert space of multiply wound heterotic strings which are dual to the collection of $I$ KK monopoles. Analysis of the spectrum and of fermionic zero modes in various duality frames indicates that in the context of this counting problem, the contribution from the tensionless strings appears not to be important.
\end{itemize}
Combining these arguments, we propose the candidate microscopic SCFT in
$\S{\ref{Four}}$ that counts the dyons that we are interested in. The resulting spectrum passes a number of nontrivial consistency checks discussed in $\S{\ref{Consistency}}$. This can be viewed as a further supporting evidence for our microscopic proposal.

\subsection{Five-dimensional black holes on $ALE$ spaces \label{Five}}

When we replace the multi-centered Taub-NUT space by the ALE space $\IC^2/\IZ_I$, we have type IIB string theory on ${ \IR \times \IC^2/\IZ_I \times S^1 \times  K3 } $. We consider a system of charges consisting of D1-branes wrapping the $S^1$, D5-branes wrapping the $K3 \times S^1$, momentum excitations $P$ along the $S^1$, and angular momentum $j$ placed at the origin of the ALE space. In $\IR \times \IC^2/\IZ_I$ we do not have Poincar\'e symmetry but can label point particle states or black holes by their mass and the spin under the unbroken symmetry group $U(1)_L \times SU(2)_R$.
Such a state is quarter-BPS, and we denote it by the five-dimensional charge vector\footnote{ We will assume that $(Q_1,Q_5)$ are relatively prime and also that they are relatively prime with respect to $n$ and $I$. We allow $(n,I)$ to have common factors.} $\G^{(5)} \equiv (Q_1,Q_5,n,j)$. In the low energy effective  $\CN=4$, $5$-dimensional supergravity description, this state is a  black hole at the center of ALE space with an entropy given by the Wald formula.  At leading order in the large charge expansion, this entropy is a quarter of the area in Planck units of the $5d$ black hole. We would like to count the entropy by counting the excitations in the microscopic description.

When $I=1$, the noncompact spacetime is $\IR^{1, 4}$, the low energy theory on the worldvolume of the branes is a  two dimensional $(4,4)$ SCFT
\be\label{theoryI=1}
X^{5d} = \s(\IR^4) \times \s({\rm Sym}^{Q_1Q_5 +1}(K3)).
\ee
The states we are interested in are purely left-moving excitations in this theory tensored with the right-moving oscillator ground state which preserves the required four supercharges. The charge $n$ is then the $L_0$ eigenvalue, and the charge $j$ is equal to the charge $l$ under the current $J_0$ which is the $U(1)_L \subset SU(2)_L$ R-symmetry current in the left moving sector.
The degeneracy  $d(\G^{(5)})$ of such states $\G^{(5)} = (Q_1,Q_5,n,l)$    is generated by an index invariant under the deformations of the theory. In the above case, the presence of the $\IR^4$ includes two complex fermion zero modes, and therefore the relevant index is a {\it modified} elliptic genus, or a helicity supertrace. \cite{Cecotti:1992qh,Gregori:1997hi,Kiritsis:1997gu}
\bea\label{defE2}
\CE_2 (X^{5d}; q,y)  \equiv  \Tr^{X^{5d}}_{RR} (-1)^{ J_0 - 
\T  J_0} \T{ J_0}^2 q^{L_0} \T q^{\bar L_0} y^{ J_0}  \equiv   \sum_{n,l} \hat{c}_1^{5d}(Q_1Q_5, n,l) q^n y^l \, .
\eea
The form of $X^{5d}$ implies that the degeneracy above only depends on the product $Q_1 Q_5$\footnote{In our notation $J_0 = 2 J_L$ and $\tilde J_o = 2 J_R$ are the R-charges of the SCFT and are integers.}.
For a product space like  (\ref{theoryI=1}) where one of the factors has two fermion zero modes and the second does not, the index $\CE_2$ is a product of the elliptic genus $\chi({\rm Sym}^{Q_1Q_5+1}(K3))$ and the modified elliptic genus of $\IR^4$:
\be\label{degI=1}
 \CE_2(\IR^4 \times {\rm Sym}^{Q_1Q_5+1}(K3);q,y)
 =  \CE_2(\IR^4; q,y) \times \chi({\rm Sym}^{Q_1Q_5+1}(K3);q,y)  .
\ee

For generic $I$, the orbifold action of $\IZ_I$ is embedded into the $U(1)_L$ of the $Spin(4) = SU(2)_L \times SU(2)_R$ symmetry of the transverse $\IR^4$ and therefore preserves the rightmoving supercharges. The orbifold action commutes with $U(1)_L$ whose quantum number we have indicated by $j$.  The low energy effective theory is thus a $(0,4)$ $2d$ SCFT. This theory can be deduced  using a quiver construction \cite{{Dabholkar:2008}, Okuyama:2005gq} generalizing the analysis of the quiver with $K3$ replaced by $T^4$. This analysis, which will be presented in detail in \cite{Dabholkar:2008}, suggests that SCFT of interest is ${\rm Sym}^I(\IR^4 \times {\rm Sym}^{Q_1 Q_5 +1}(K3))/\IZ_I$ where the $\IZ_I$ belongs to the $SU(2)_L$ R-symmetry of the parent theory. It acts on the left moving fermions and breaks all the supersymmetry generated by them. It also breaks the $SU(2)_L$ symmetry to a $U(1)$ subgroup. This surviving left-moving $U(1)$ symmetry of the orbifold theory is no longer an $R$-symmetry but can still be used to define the quantum number $j$.

The orbifold action generates a new $\IZ_I$ quantum symmetry which labels the twist sectors. The black hole state we are interested in carries zero charge under this symmetry, and so can focus our attention to the untwisted sector of the orbifold, {\it i.e.} the invariant states in the parent $(4,4)$ theory. Furthermore, because of the $\IZ_I$ projection, a state with $U(1)_L$ charge $j$ in the orbifold theory descends from a state with $J_0$ R-charge $l=Ij$ in the parent $(4,4)$ theory.
Therefore, the degeneracy of states of the form $\G^{(5)} = (Q_1, Q_5, n,l;I)$  in the $ALE$ space are given by
\begin{equation}\label{5ddegen}
   d^{5d}_I(Q_1,Q_5,n,j) =  \hat{c}^{5d}_I(Q_1Q_5,n,l=Ij) \, ,
\end{equation}
where $\hat c$ is defined by the Fourier expansion  of the modified elliptic genus
\bea\label{theoryI}
\CE_2({\rm Sym}^I(X^{5d})) =
\sum_{n,l} \hat{c}^{5d}_I(Q_1Q_5,n,l) q^n y^l .
\eea

Whenever the sigma model on $X^{5d}$ has two fermion zero modes, one can show that  the contributions to the quantity $\CE_2(X^{5d})$ only arise from each Hilbert subspace of strings of length $r$ which divides $I$, and in each such Hilbert space, we count excitations of integer momentum $nr/I$ and $J$ charge $lr/I$ \cite{Maldacena:1999bp}. We review this theorem in $\S{\ref{theorem}}$. Using this theorem, we can now write the degeneracy of the symmetrized theory in terms of the degeneracy of dyons in flat space (\ref{degI=1}):
\bea\label{deg5d}
\hat{c}^{5d}_I(Q_1Q_5,n,l) & = & \sum_{r|I, I|rn, I|rl} {\frac{I}{r}} \; \hat{c}^{5d}_1(Q_1Q_5, \frac{nr}{I} r, \frac{l r}{I}) \cr
& = & \sum_{s|I, s|n, s|l} s \; \hat{c}^{5d}_1(Q_1Q_5, \frac{n I}{s^2}, \frac{l}{s})
\eea

For large charges, the leading contribution to the above degeneracy comes from the $s=1$ term with the asymptotics
\be\label{deg5dlead}
d^{5d}_I(\G^{(5)}) \sim \exp\left[ 2\pi\sqrt{Q_1 Q_5 n I - I^2 j^2/4} \right],
\ee
which is indeed in agreement with the Bekenstein-Hawking entropy of the $5d$ black hole  in ALE space \cite{Gaiotto:2005gf}. The shift of $l^2/4 = I^2j^2/4$ can be understood as coming from the spectral flow in the parent $(4,4)$ theory.
The correctness of the subleading terms cannot be checked independently for the $K3$ case. But a similar analysis for $T^4$ results in answers consistent with large $E_{7, 7}(\mathbb{Z})$ duality invariance in that case giving additional evidence for the proposed SCFT \cite{Dabholkar:2008}.

\subsection{Four-dimensional black holes \label{Four}}

The answer for the five-dimensional in (\ref{deg5d}) is highly suggestive. For example, the first term with $s=1$ comes just from the elliptic genus for symmetric product of $K3$ which is closely related to the Igusa cusp form which determines the four-dimensional degeneracies. The Igusa cusp form differs from the elliptic genus of the symmetric product by a multiplicative piece which was called `Hodge anomaly' in \cite{gritsenko-1999-}. This piece is crucial for the automorphic properties of the Igusa form under $Sp(2, \IZ)$. Physically, it  corresponds to the contribution of KK-P bound states \cite{Dabholkar:2006xa, David:2006yn} and is essential for obtaining an S-duality invariant spectrum. The Hodge anomaly is however different for each piece in the sum which suggests that these contributions also have to be included appropriately in the symmetrization. Below we will argue for the same from a microscopic point of view.

To understand the entropy of $4d$ black holes, we go back to the situation when the radius $R$ of the charge $I$ Taub-NUT space is large but finite. The spacetime is $\IR \times$ $TN_I \times S^1 \times K3 $ with the branes wrapping $S^1$ and $K3$ as before. The presence of the KK monopole, even for $I=1$, breaks the $SU(2)_L$ symmetry to $U(1)_L$. The low energy effective theory is again governed by a $(0,4)$ SCFT in two dimensions living on $\IR \times S^1$. 

The case of $I=1$ KK monopole was analyzed in \cite{{Shih:2005uc}, David:2006yn}. We can recast the computation as counting BPS states in the sigma model 
\be\label{theory4dI=1}
X^{4d} = \s(\mathrm{Sym}^{Q_1Q_5+1}(K3)) \times \s(TN_1) \times \s_L(\textrm{KK-P}) \, .
\ee
The first factor is the same $(4,4)$ theory that entered the $5d$ computation. The piece $\s(TN_1)$ describes the bound states of the center of mass of the $D1$-$D5$ with the KK monopole, it is a $(0,4)$ theory with two fermion zero modes. The piece $\s_L(\textrm{KK-P})$ describes the bound states of the KK monopole and momentum and is  a  conformal field theory of $24$ left-moving bosons of the heterotic string, which can be deduced from the duality between the Type-IIB KK-P system  and the heterotic F1-P system. We briefly discuss the latter two pieces in $\S{\ref{E2Comp}}$.

The system  of $I$ KK-monopoles contains many bound states at threshold. Analysis of the bound state problem directly from the quantization of the moduli space of KK-monopoles is quite subtle in general. However, we can deduce the structure of the bound states from the dual heterotic string.  BPS states with KK-monopole charge $I$ in the Type-IIB frame map to BPS states with fundamental string winding charge $w= I$ in the heterotic frame. From the analysis of the perturbative Hilbert space of the heterotic string, we know that this sector will contain single particle states of multiply wound strings  as well as multi-particle states of singly wound strings and various possible combinations in between. 
The important point though is that most multiparticle states will have far too many fermionic zero modes to contribute to the index of our interest.  For our purposes we need to focus our attention only on those states that have exactly two complex fermionic zero modes. 

It is illuminating to study the dual F1-P system in the heterotic frame to gain further insight. As explained in detail in $\S{\ref{Multiple}}$, the only contributions come from properly symmetrized wavefunctions of a collection of  KK-monopoles all of charge $s$,  for all $s$ which divide $I$.

Consider now the full problem of ${TN}_I$ and D1-D5 with momentum excitations. The D-branes can be thought of as internal quantum numbers to the effective theory on the string describing the $I$ monopoles and must be symmetrized along with it. As before, the string must divide equally into the shorter strings of equal length. When momentum is added, the only contribution comes when it has a common factor with $I$.

Note that this procedure preserves the four dimensional electric-magnetic duality of the system which is nontrivial. It turns out that the same procedure indeed gives the correct answer for the $\CN=8$ theory on the $T^4$ which can be independently checked using the larger duality group in that case \cite{Dabholkar:2008}.

Let us note that the dyon is free to move in the transverse $\IR^3$. We are interested in a black hole sitting at rest and hence do not integrate over the bosonic zero modes corresponding to this motion. Hence we also do not integrate over the four complex fermionic partners of these center of mass coordinates. These coordinates are common to all subsystems and their quantization gives rise to an overall multiplicative factor for the degeneracy.

The precise proposal is now that the theory describing the low energy excitations of our system of interest is the $(0,4)$ theory $\mathrm{Sym}^I(X^{4d})/\IZ_I$. The states considered in (\ref{final charges}) have no winding charge around the Taub-NUT which correspond to the quantum symmetry in the orbifold limit. As before, because of the $\IZ_I$ projection, a state with $U(1)_L$ charge $\t n$ in the orbifold theory descends from a state with $J_0$ R-charge $l=I \t n$ in the parent $(4,4)$ theory.
Therefore, the degeneracy of these states is given by\footnote{The factor of $(-1)^{l}$ noted in \cite{Shih:2005he,{Cheng:2007ch}} arises because the charge quantum number $\t n$ in four dimensions becomes the spin quantum number $j$ in five dimensions. The overall negative sign arises because we are counting the number of bosonic supermultiplets minus the number of fermionic supermultiplets \cite{Sen:2007qy}.}
\begin{equation}\label{4ddegen}
   d^{4d}_I(Q_1,Q_5,n,\t n;I) =  (-1)^{l+1} \; \hat{c}^{4d}_I(Q_1Q_5,n,l=I \t n;I) \, ,
\end{equation}
where the $\hat c$ are defined by the Fourier expansion  of the modified elliptic genus   of the $(4,4)$ theory $\mathrm{Sym}^I(X^{4d})$.
\be\label{sm4dI}
\CE_2(\mathrm{Sym}^I(X^{4d})) \equiv \sum_{N,l} {\hat c}^{4d}(Q_1,Q_5,1,n,l;I) q^n y^l
\ee
Following the same argument as in the previous section and using the theorem in $\S{\ref{theorem}}$, we find
\be\label{deg4d}
\hat{c}^{4d}(Q_1,Q_5,n,l;I)  =  \sum_{s|I, s|n, s|l} s \; \hat{c}^{4d}(Q_1,Q_5, \frac{n I}{s^2}, \frac{l}{s};1)
\ee
This degeneracy is given in terms of the modular form $\Phi_{10}$ and we can express the answer as:
\bea\label{deg4dform}
\hat{c}^{4d}(Q_1,Q_5,n,l;I)  & = & \sum_{s|I, s|n, s|l} s \int_{\CC} d\s d\rho d v \; \exp\{- \pi i (\rho \frac{n I}{s^2} + \s Q_1Q_5 + 2 v
\frac{l}{s})\} \frac{1}{\Phi_{10}(\r,\s,v)}    \cr
& = & \sum_{s|I, s|n, s|l} s \oint_{\CC_s} d\s d\rho d v \, s^3 \, \exp\{- \pi i (\rho n I + \s Q_1Q_5 + 2 v l )\} 
\frac{1}{\Phi_{10}(s^2 \r,\s,s v)}  \cr
& = & \sum_{s|I} s \oint_{\CC} d\s d\rho d v \, 
\exp\{- \pi i (\rho nI + \s Q_1Q_5 + 2 v l )\}   \frac{1}{\Phi_{10}(s^2 \r,\s,s v)}
\eea
The contours $\CC_{s}$ above are defined  by 
\bea\label{contours}
& 0 \le \Re(\s) \le 1, \quad 0 \le \Re(\rho) \le \frac{1}{s^2}, \quad 0  \le \Re(v) \le \frac{1}{ s}\cr
 & {\Im(\rho)} >>1, \quad \Im(\sigma) >>1, \quad \Im(v )>> 1,
\eea
with $\CC \equiv \CC_{1}$.  In terms of the invariants $(Q^2, P^2, Q.P)$ and  $I=\gcd(Q \wedge P)$ of the charge vector $\G$ (\ref{final charges}), we can rewrite this formula as
\begin{equation}\label{finalomega}
    \Omega_I (\G) = (-1)^{Q.P+1}  \oint_{\CC} d\s d\rho d v \, \exp{\left(-i\pi \, \Gamma^{t} \cdot
      \left(\begin{array}{cc}
                \rho & v \\
                v & \sigma \\
              \end{array}
            \right)\cdot \Gamma \right) \, {\mathcal{Z}_I(\tau)} }\, ,
\end{equation}
with
\begin{equation}\label{finalpart}
    {\mathcal{Z}_I(\r, \sigma, v)} = \sum_{s|I} s \, \frac{1}{\Phi_{10}(s^2 \r,\s,s v)} \, .
\end{equation}
The degeneracy can also be expressed as
\begin{equation}\label{finalomega2}
    \Omega_I\left(\frac{Q^2}{2}, \frac{P^2}{2}, Q\cdot P \right) =  \sum_{s|I} s \, \Omega_1\left(\frac{Q^2}{2s^2}, \frac{P^2}{2}, \frac{Q\cdot P}{s} \right).
\end{equation}

\section{Consistency checks \label{Consistency}}

The degeneracy of dyons $\Omega(\Gamma, \phi)$ that follows from the proposed microscopic SCFT must satisfy a number of physical requirements.
\vspace{-3mm}
\begin{myenumerate}
  \item It must be integral and  S-duality invariant.
  \item For large charges, $\log (\Omega(\Gamma, \phi))$ must agree with Wald entropy of corresponding black holes.
  \item For small charges, it must agree with the degeneracy of corresponding field theory dyons.
  \item It must display the correct moduli dependence and walls of marginal stability.
\end{myenumerate}

Indeed, imposing all these four requirements and from the experience with the $I =1$ case, a dyon partition function was proposed in \cite{{Banerjee:2007sr}, Banerjee:2008ri} which exactly coincides with the partition function (\ref{finalpart}) that we have derived from the microscopic SCFT. In the macroscopic analysis of  \cite{{Banerjee:2007sr}, Banerjee:2008ri},  the form of the various terms in the partition function and their relative coefficients were fixed  by hand by demanding the first three of these requirements. The resulting formula then satisfied the fourth requirement in a nontrivial way.
Our microscopic derivation automatically gives the correct  structure as well as  the relative coefficients of all terms in the partition function in a very natural way. 

It is satisfying that the  modified elliptic genus of the microscopic SCFT  naturally incorporates all macroscopic physical requirements.  We briefly discuss below some of the salient points of these macroscopic tests, which further support the arguments that we have used to deduce the microscopic theory.

\subsection{Duality properties and moduli dependence\label{Duality2}}

Integrality of the degeneracy for $I >1$  follows from the integrality of the degeneracy for the $I=1$ case from (\ref{finalomega2}).

To check for S-duality invariance, let us see how the physical S-duality group $SL(2,\IZ)$ is embedded in the $Sp(2, \mathbb{Z})$. This can be deduced from our experience with the case of $I=1$ by demanding S-duality invariance of the spectrum. Consider an S-duality transformation $\Gamma \rightarrow \Gamma^\prime = h \Gamma$
\begin{equation}\label{stranform}
   \Gamma  = \left[
               \begin{array}{c}
                 Q \\
                 P \\
               \end{array}
             \right] \rightarrow
             \left[
               \begin{array}{c}
                 Q^\prime \\
                 P^\prime \\
               \end{array}
             \right] =
             h \left[
               \begin{array}{c}
                 Q \\
                 P \\
               \end{array}
             \right]
   \qquad h =
   \left(
     \begin{array}{cc}
       a & b \\
       c & d \\
     \end{array}
   \right)
   \in SL(2, \IZ)\, .
\end{equation}
under which the moduli $\phi$  transform to $\phi'$ taking us to a new chamber $\bf X'$. In particular, the axion-dilaton modulus transforms as
\begin{equation}\label{axion dilaton}
    S \rightarrow S'= \frac{a S + b}{cS +d},
\end{equation}
and other moduli are fixed. The degeneracy for the new set of charges in the new chamber is given by
\begin{equation}\label{prime}
    \Omega_I(\Gamma^\prime, \phi^\prime) = \int_{\CC'}
   d^3 \tau' \, e^{-i\pi {\Gamma^\prime}^{t} \cdot
     \tau' \cdot \Gamma^\prime}\,
   {\CZ_I(\tau^\prime)} \, .
\end{equation}
To exhibit S-duality invariance, we demand that ${\Gamma'}^t \tau'\cdot \Gamma' = \Gamma^{t} \cdot\tau \cdot \Gamma$, that is, $ \tau^\prime = (h^t)^{-1} \tau h^{-1}$,
Now, such a transformation of $\tau$  can be viewed as an $Sp(2, \mathbb{Z})$ transformation
\begin{equation}\label{sembed}
\left(
  \begin{array}{cc}
    A & B \\
    C & D \\
  \end{array}
\right)
   = \left(
      \begin{array}{cc}
        (h^t)^{-1} & \textbf{0} \\
        \textbf{0} & h   \\
      \end{array}
    \right)    =
    \left(
      \begin{array}{cccc}
        d & -c & 0 & 0 \\
        -b & a & 0 & 0 \\
        0 & 0 & a & b \\
        0 & 0 & c & d \\
      \end{array}
    \right)
   \, \in  Sp(2, \mathbb{Z}).
\end{equation}
This defines the embedding of the physical duality group into the $Sp(2, \mathbb{Z})$. The partition function $\CZ_I$ is however not invariant under the full $SL(2, \IZ)$ but only under a congruence subgroup
$\Gamma^0(I)$ defined by matrices 
\begin{equation}\label{definition}
     h =
   \left(
     \begin{array}{cc}
       a & b \\
       c & d \\
     \end{array}
   \right)
   \in SL(2, \IZ)\,  \quad b = 0\,\, \mod\,  I \,\, .
\end{equation}
To see the invariance of the partition function  under this subgroup, we consider the transformation properties under each of the terms (\ref{finalpart}) under $h$. Note that
\bea
\left(\begin{array}{cc}
     s^2\rho' & sv' \\
     sv' & \sigma' \\
\end{array}\right)
=
\left(\begin{array}{cc}
     d & -sc \\
     -b/s & a \\
\end{array}\right)
\left(\begin{array}{cc}
     s^2\rho & sv \\
     sv & \sigma \\
\end{array}\right)
\left(\begin{array}{cc}
     d & -b/s \\
     -sc & a \\
\end{array}\right)
\eea
Using the fact that $b=0 \, \mod \, (I)$ and using the modular properties of $\Phi_{10}(\rho,\sigma,v)$ under $Sp(2, \mathbb{Z})$, it is easy to see that
\bea
 \Phi_{10}(s^2\rho',\sigma',sv')=\Phi_{10}(s^2\rho,\sigma,sv).
\eea
Hence $\CZ_I$ defined by (\ref{finalpart}) is invariant under $\Gamma^0(I) \subset Sp(2, \IZ)$ transformations of the form (\ref{sembed}) with $ b = 0\,\, \mod\,\, I$.

To prove duality invariance under $\Gamma^0(I)$, we can change the integration variable from $\tau$ to $\tau'$.
Using the above transformation properties we see that
\begin{eqnarray}
   d^3 \tau' &=& d^3 \tau \, ,\\
  \mathcal{Z}_I ( \tau') &=& \mathcal{Z}_I(\tau)\, , \\
  \Gamma^{\prime t} \cdot
     \tau'\cdot \Gamma' &=& \Gamma^{t} \cdot
     \tau \cdot \Gamma.
\end{eqnarray}
Under these change of variables the integration contour $\CC'$ goes to $\CC$ and we obtain
\begin{equation}\label{sinvariance}
    \Omega_I(\Gamma^\prime, \phi^\prime) = \Omega_I(\Gamma,  \phi) \, .
\end{equation}

We therefore see that the degeneracy defined by (\ref{finalomega}) is manifestly invariant under $\Gamma^0(I)$. The complete spectrum of dyons should of course be invariant under the full $SL(2, \IZ)$ which is the physical duality group. To exhibit this duality invariance we proceed as follows. Consider the coset
\begin{equation}\label{coset}
    SL(2, \IZ)/\Gamma^0(I) \equiv \{g_0, g_1, \ldots g_k\} \, , 
\end{equation}
where $k$ is the order of $\Gamma^0(I)$ in $SL(2, \IZ)$ and $g_0$ is the identity element. The function $\CZ_I$ is not invariant under the full $SL(2, \IZ)$. Under the action of an element $g_l$ it will transform to a  new function which we denote  by $\CZ^{(l)}_I$ with the convention that $\CZ^{(0)}_I = \CZ_I$.The charge vector $\G$ of the dyon will also transform under this element to a new charge vector $\G^{(l)}$. The invariants of this transformed charge vector will no longer be of the form $(r_1,r_2,r_3,u_1) = (I,1,1,1)$. However, the degeneracy of these dyons $\G^{(l)}$ can be defined simply by using the transformed partition function $\CZ^{(l)}_I$. With these definitions, the resulting spectrum is thus manifestly duality invariant under the full duality group $SL(2, \IZ)$.

The formula for the degeneracy defined in (\ref{inverse3}) does not appear to have any moduli dependence but it is contained secretly in the choice of the contour \cite{{Sen:2007vb},{Dabholkar:2007vk},Cheng:2007ch}.  Since the degeneracy is really an index, it does not have continuous moduli dependence. However, it can have discrete jumps while crossing walls of marginal stability where the state is marginally stable to decay into half-BPS states. This discrete dependence on the moduli is nicely encoded in the choice of the contour. The walls  divide up the moduli space into chambers $(\textbf{X}, \textbf{X}^\prime, \ldots)$. The formula above, defined using a contour $\CC$ is valid for $\phi$ belonging to a specific chamber $\textbf{X}$. Different contours  $(\CC, \CC', \ldots )$ which cannot be deformed into each other without crossing poles of the partition function are in one-to-one correspondence with the chambers $(\textbf{X}, \textbf{X}^\prime, \ldots)$. Crossing a pole in the $\tau$ plane corresponds to crossing a wall in the moduli space. Hence, the difference in the degeneracies in going from a chamber $\bf X$ to a chamber $\bf X^\prime$ by crossing a  wall  is simply given by the residue at the pole that is crossed while deforming $\CC$ to $ \CC^\prime$. The pole structure of the partition function thus captures the structure of the walls  in a beautifully consistent manner for all values of $I$. Precise relation between the contours and the chambers for $I=1$ can be found in \cite{{Sen:2007vb}, Cheng:2007ch}. Analysis of walls for $I>1$ can be found in \cite{Banerjee:2008ri,{Banerjee:2008pu}}.

\subsection{Comparison with black holes and  field theory dyons \label{Comparison}}

In the large charge limit, the  degeneracy $\Omega(\Gamma, \phi)$ derived above can be compared with  degeneracy of corresponding dyonic black holes. More precisely, we take the invariant $\Delta$ to be very large, $\Delta >>1$, and  can take  $Q^2$, $P^2$, $Q\cdot P$ all large and of the same order.  The leading asymptotics to the degeneracy is determined by the first term in (\ref{finalpart}) with $s=1$ for all values of $I$. This term is identical to  the $I=1$ case with the same value of $\Delta$, and hence the resulting statistical entropy matches with thermodynamic Wald entropy including the  subleading correction  \cite{{LopesCardoso:2004xf}, LopesCardoso:2006bg} that is suppressed by inverse powers of charges, coming for four derivative corrections to the Einstein-Hilbert action.

The black hole entropy is not sensitive to the terms in (\ref{finalpart}) corresponding to $s >1$ which are exponentially suppressed compared to the $s=1$ term. To see their contribution, we must consider the opposite limit of small charges. In this limit, we consider the dyons with
\begin{equation}\label{field}
    \frac{Q^2}{2} = -I^2, \quad \frac{P^2}{2} =-1, \quad Q\cdot P = I.
\end{equation}
These dyons correspond to  Stern-Yi dyons in $SU(3)$ gauge theories which arises as a field theory limit of string theory as explained in \cite{{Sen:2007ri}, Dabholkar:2008tm}. The field theoretic degeneracy of the Stern-Yi dyons has been computed independently using various methods \cite{{Stern:2000ie}, Denef:2002ru, Dabholkar:2008tm}, and  is known to equal $I$. This precisely matches  predictions from the string-theoretic partition function that we have derived from the microscopic SCFT. The factor of $I$ comes from the term $s=I$ in (\ref{finalpart}) and hence is sensitive to the exponentially subleading terms.

We thus see that the  degeneracy that follows from our microscopic partition function is consistent with the physical expectations in the two opposite limits of  large and small charges.

\section{Conclusions \label{Conclusions}}

We have shown that the partition function for dyons for all values of $I$ can be derived from a modified elliptic genus of a microscopic $(0, 4)$ SCFT in a uniform way. The resulting degeneracies satisfy a number of nontrivial physical consistency checks. This derivation relied on a specific representative in the duality orbits. Since $I$ is the unique discrete duality invariant relevant to this problem, one can deduce the degeneracies of \text{all} dyons that lie in the duality orbits of these representative ones and then extend them appropriately as explained in $S{\ref{Duality2}}$ to obtain a dyon spectrum that is invariant under the full duality group $G(\mathbb{Z})$.

We should emphasize that we have used a specific index in the microscopic theory which is proportional to the helicity supertrace $B_6 = Tr (-1)^{2\lambda} (2\lambda)^6$ in spacetime where $\lambda$ is the 4-dimensional helicity.
Therefore,  we have counted the number of bosonic minus fermionic supermultiplets of  \textit{all} dyons that come in 64-dimensional super-multiplets and which do not have any additional fermionic zero modes. We have seen that the structure of the fermionic zero modes plays a very crucial role in the derivation. The analysis of the more familiar case of perturbative multiparticle Hilbert space of multiply wound strings in $\S{\ref{FP}}$ shows that many configurations which  do exist in the spectrum may not contribute to this particular index if they have additional fermionic zero modes. Indeed, the analysis of field theoretic dyons indicates, that such dyons must be present which correspond to string networks that are not planar \cite{{Bergman:1997yw}, Bergman:1998gs}. Such dyons  with additional faces are expected also from the analysis of \cite{Dabholkar:2006bj} which correspond to  higher genus worldsheets of M5-branes. Generically these dyons will have far many more zero modes. It would be interesting to see if an index or a similar quantity other than $\CE_2$ can be used to count these dyons.

This discussion of fermionic zero modes is relevant also for seeing why the entropy enigma noted is \cite{Denef:2007vg} in the context of $\CN=2$ dyons is not relevant in the present context. Multi-centered black holes which can have more entropy than the single-centered black holes can exist also in $\CN=4$ supergravity. However, if they have more fermionic zero modes, then they will not contribute to the index $B_6$ and hence the index will count only the  single-centered black holes.

By replacing $K3$ by $T^4$, we can perform a similar analysis of dyon spectrum in the resulting $\CN=8$ theory. In this case, the duality group is bigger, and as a result tighter duality constraints are possible which can be used to learn more about the microscopic SCFT. The details will be presented in a forthcoming publication \cite{Dabholkar:2008}.

\subsection*{Acknowledgments}

It is a pleasure to thank  Suresh Nampuri, Boris Pioline, Greg Moore, and Ashoke Sen for valuable discussions. The work of A.~D. was supported in part by the Excellence Chair of the Agence Nationale de la Recherche (ANR). 
The work of J.~G. was supported in part by Funda\c{c}\~{a}o para Ci\^{e}ncia e Tecnologia (FCT). S.~M. would like to thank TIFR and HRI for hospitality where part of this work was completed.

\appendix

\section{Appendices}

\subsection{The structure of the modified elliptic genus $\CE_2$ \label{theorem}}

Consider a SCFT with target space $X$ which has complex fermionic zero modes.
The elliptic genus of such a theory vanishes due to the presence of these zero modes.
As was shown in \cite{Cecotti:1992qh,Gregori:1997hi,Kiritsis:1997gu}, there are other topological indices, which do not vanish for such theories. These are defined by adding insertions of the fermion number operator into the elliptic genus which soak up the above-mentioned zero modes.

When the theory $X$ has precisely two fermionic zero modes, the non-vanishing index is the \textit{modified
} elliptic genus of a SCFT on $X$:
\be
\CE_2(X;q,\bar{q},y)=\mathrm{Tr}_{\mathrm{RR}}(-1)^{J_{0}-\tilde{J}_{0}}(\tilde{J}_{0})^{2}q^{L_{0}}\bar{q}^{\tilde{L}_{0}},\ee
where $J_{0}$ and $\tilde{J}_{0}$ are the left and right
R-charges.
The presence of the $(J_0^3)^2$ soaks up the fermionic zero modes and give a non-vanishing answer.
The index $\CE_2$ is invariant under smooth deformations of $X$ because the massive representations {\it i.e.} the long representations of the $\mathcal{N}=4$ algebra do not contribute\footnote{There are subtleties which arise from additional charges {\it e.g.} in the case $X=T^4$, from winding and momentum modes \cite{Maldacena:1999bp}.}. It's easy to see
that for states with $L_{0}>0$ we have $\rm{Tr}_{j}(-1)^{J_{0}}=\rm{Tr}_{j}(-1)^{J_{0}}J_{0}=\rm{Tr}_{j}(-1)^{J_{0}}(J_{0})^{2}=0$.

In the case that the SCFT is a symmetric product $Y = \mathrm{Sym}^n(X)$, the modified elliptic genus has a special structure -- it only gets contributions from some parts of the Hilbert space. In the long string interpretation of the symmetric product Hilbert space \cite{Dijkgraaf:1996xw}, only the strings of lengths which are {\it divisors} of the length of the longest string contribute. This is in contrast with the case when $X$ has no fermionic zero modes, when the strings of {\it all} lengths smaller than the longest one contributes. We review this theorem in this appendix.

We start by considering
the partition function
\be
Z(X;q,\bar{q},y,\bar{y})=
\mathrm{Tr}_{\mathrm{RR}}(-1)^{J_{0}-
\tilde{J}_{0}}q^{L_{0}}\bar{q}^{\tilde{L}_{0}}y^{J_{0}}
\tilde{y}^{\tilde{J}_{0}}.
\ee
and then we take its second derivative
\be
\partial_{\tilde{y}}^{2}Z\mid_{\tilde{y}=1}=\CE_2.
\ee
Let the Fourier expansion of this function be
\begin{equation}\label{fourierZ}
Z(X)=\sum_{\Delta,\bar{\Delta},l,\bar{l}}
c(\Delta,\bar{\Delta},l,\bar{l})q^{\Delta}\bar{q}^{\bar{\Delta}}y^{l}\tilde{y}^{\bar{l}} \, .
\end{equation}
To compute the modified elliptic genus of a symmetric product we first
determine the generating function
\be
\textbf{Z}(p,q,\bar{q},y,\tilde{y})=
\sum_{N=0}^{\infty}p^{N}Z(\mathrm{Sym}^{N}(X))=
\prod_{n=1}^{\infty}\prod_{\Delta,\bar{\Delta},l,\bar{l} }
\frac{1}{\left(1-p^{n}
q^{\frac{\Delta}{n}}\bar{q}^{\frac{\bar{\Delta}}{n}}y^{l}
\tilde{y}^{\bar{l}}\right)^{c(\Delta,\bar{\Delta},l,\bar{l})}},
\ee
where the sum is over weights that satisfy ${\frac{\Delta -\bar \Delta}{n} \in \IZ}$.
The invariance of the $\CE_2$ under smooth deformations
of $X$ translates into the condition
\be
\sum_{\bar{l}}\bar{l}^{2}c(\Delta,\bar{\Delta},l,\bar{l})=0\,\,\,\,\,\mathrm{for}\ \ \bar{\Delta}>0,
\ee
and the existence of two complex fermion zero modes implies $Z(X)\mid_{\tilde{y}=1}=Z(X)'\mid_{\tilde{y}=1}=0$,
or
\begin{eqnarray*}
\sum_{\bar{l}}c(\Delta,\bar{\Delta},l,\bar{l}) & = & 0,\\
\sum_{\bar{l}}\bar{l}c(\Delta,\bar{\Delta},l,\bar{l}) & = & 0,
\end{eqnarray*}
respectively. Proceding with two derivatives and taking into account the last constraints we find
\be\label{modell}
\frac{1}{2}\partial_{\tilde{y}}^{2}
\textbf{Z}\mid_{\tilde{y}=1}=\sum_{s,n,m,l}s\left(p^{n}q^{m}y^{l}\right)^{s}\hat{c}(nm,l),
\ee
where $\hat{c}(\Delta,l)\equiv\frac{1}{2}\sum_{\bar{l}}\bar{l}^{2}c(\Delta,0,l,\bar{l})$.
If we want $\CE_2(\rm{Sym}^{I}(X))$ we have to pick the
coefficient of $p^{I}$ in the above formula, which gives us
\be
\CE_2(\mathrm{Sym}^{I}(X);q,y)=\sum_{s\mid I, s\mid N, s\mid L}sq^{N}y^{L}\hat{c}(\frac{IN}{s^2},\frac{L}{s}).
\ee

This formula shows that the full Hilbert space separates into a
sum over subspaces graded by the divisors of $I$. Lets see more carefully
what is happening in the long string interpretation. The structure of the Hilbert space is of the form:
\be\label{hilbspstr}
\mathcal{H}(\mathrm{Sym}^{I}(X))=\oplus_{\{s_{r}\}}\otimes_{r>0}\mathrm{Sym}^{s_{r}}(\mathcal{H}_{r}(X)),
\ee
where the sum is over the partitions $\sum_{r<I}rs_{r}=I$ {\it i.e.},
over the conjugacy classes of the symmetric group of $I$ objects. We denote the Hilbert space of a single ``string'' of length $r$ as $\CH_r(X)$. In each $\CH_r$  only momenta which are compatible with the length of the string are allowed. In the orbifold language, this is simply saying that each $\CH_r$ is a $\IZ_r$ invariant space, $\IZ_r$ being the centralizer subgroup in effect.

The partition function corresponding to each term in the above direct sum (\ref{hilbspstr}) is a product of partition functions, one for each length $r$. When we apply two derivatives, all  terms with more than one $r$ in the product will vanish. The only surviving sectors are the ones with a single $r$. This corresponds to partitions of $k$  which are integer factorizations $rs_{r}=k$ which physically  correspond to symmetrized subspaces composed of  strings of equal length. Because of the symmetrization, they preserve exactly two fermion zero modes. The overall factor of $s$ in each such subspace can be understood as arising from the two derivatives applied to each factor, which has to do with the specific index $\CE_2$ that we are computing.

\subsection{Counting states in $\s(\IR^4)$, $\s(TN)$, and $\s({\rm KK-P})$ \label{E2Comp}}

The sigma models $\s(TN)$, $\s(\textrm{KK-P})$ and $\s(\IR^4)$  contribute BPS states to the $4d$ and $5d$ counting problems discussed in the main text. In this appendix, we shall briefly sketch the counting of BPS states in these models. In each of these models, the relevant states are quarter-BPS or half-BPS and captured by different number of insertions of the fermion number operator.

In the $5d$ counting problem, the model $\s(\IR^4)$ arises as the space transverse to the D1-D5 brane system describing the motion of the center of mass. It consists of two free complex bosons and fermions which transform as a vector multiplet \cite{Maldacena:1999bp}.
This is to say that the rotations of the $\IR^4$ acts as an $R$-symmetry in the $2d$ SCFT and so the bosons are also charged under it.
The two complex fermion zero modes mean that the first non-vanishing index is:
\bea\label{E2R4}
\CE_2(q,y) & = & 4 (y^\half - y^{- \half})^2 \prod_{n=1}^\infty
\frac{(1 - q^n)^4 }{(1-q^ny)^2(1-q^ny^{-1})^2 }  \cr
& = & -4 (y^\half - y^{- \half})^4
\frac{\eta^6(q)}{ \vartheta_1(q,y)^2}.
\eea
Note that the bosons also have zero modes which will give rise to an infinite volume which multiplies the partition function. However, the translational invariance of the system allows us to divide this factor out, and the above partition function is really per unit volume.

In the $4d$ counting problem, the space transverse to the brane system is the Taub-NUT geometry with charge one, and the sigma model describing the motion of the center of mass is $\s(TN)$. When the radius $R$ of the Taub-NUT geometry is large compared to string scale, the space looks almost like $\IR^4$. However, in this case there is no translational invariance and we should compute the full partition function including the volume factor. 
There are no left moving fermion zero modes in the Taub-NUT space, but there are still two complex fermionic zero modes on the rightmoving side. Thus we should compute the modified index $\CE_2$ \cite{David:2006yn}:
\be\label{E2TN}
\CE_2(q,y)  =   4  \frac{\eta^6(q)}{\vartheta_1(q,y)^2}.
\ee

The sigma model $\s(\textrm{KK-P})$ describes the momentum excitations of the KK monopole. This system is dual to the fundamental heterotic string with leftmoving momentum excitations. It has four complex rightmoving fermionic zero modes, and we need as many insertions of the fermion number operator. The number of half BPS states is counted by the partition function:
\be\label{heterotic}
Z(q)  =   16  \frac{1}{\eta^{24}(q)}.
\ee

\subsection{Multiparticle Hilbert space of the F1-P system \label{FP}}

We will try to understand aspects of the  bound states of multiple KK-monopoles and momentum from the dual heterotic string theory on $T^4 \times S^1 \times \tilde S^1 \times \mathbb{R}^{1,3}$. In the heterotic frame we have an F1-P system of BPS states  with winding $w=I$ and momentum $n$ \cite{{Dabholkar:1989jt}, Dabholkar:1990yf}. Some aspects of the bound states of multiple KK monopoles and momentum have been analyzed also in \cite{Srivastava:2006xn}.  What we describe below is essentially a physical restatement of (\ref{modell}) which is illuminating for understanding the physics of the KK-momentum bound states and the associated fermionic zero modes.
 
A KK-monopole of unit charge in the Type-IIB frame is dual to a fundamental heterotic string winding once around the circle $S^1$. The effective  worldvolume theory of the KK-monopole on $\mathbb{R} \times S^1$ is therefore exactly that of a heterotic string in static gauge. The left-moving theory has $24$ left-moving bosons. The right-moving theory is the Green-Schwarz superstring with $8$ bosons and $8$ fermions. This can be derived also directly from the analysis of various massless modes of Type-IIB on $\mathbb{R}^{1} \times TN_1  \times S^1 \times   K_3 $ \cite{Sen:2007qy}. The worldvolume theory is a  $(0, 4)$ superconformal field theory.

For KK-monopole charge $\tilde K = I $ the situation is more complicated and one has to worry about various bound states at threshold in multiple KK-monopole moduli space. Duality with the fundamental string is very useful here to deduce the structure of these bound states.  On the heterotic side, we have to look for all states with total winding number $w =I$. We know at weak coupling that a single string wound $I$ times exists as a BPS state in the single-particle Hilbert space and hence in the dual IIB description we expect that a bound state of threshold should exist with total charge $\tilde K = I$ in the multiple KK-monopole moduli space. To count all states, we have to keep track of all such bound states at threshold.

Not all states with the correct quantum numbers need contribute to the index under consideration. The index will vanish for all states that have an excess of zero modes. To see this more clearly, it is instructive to analyze in some detail  the structure of the perturbative Hilbert space of BPS strings. We will work at zero string coupling so the theory is free.  A generic configuration consists of $n_r$ strings of winding number $r$ such that $\sum_{r=1}^w r n_r = w$. When some of the strings are identical, one has to symmetrize their wavefunctions. Together, the Hilbert space can be written as
\be\label{genHilb}
\mathcal{H}  =  \oplus_{\{n_{r}\}}\otimes_{r>0}\mathrm{Sym}^{n_{r}}(\mathcal{H}_{r}(X)), \quad \sum_{r=1}^w r n_r = w
\ee
Not surprisingly, this is nothing but the Hilbert space of the symmetric product of $w$ singly wound strings $ \mathcal{H}(\mathrm{Sym}^{w}(X))$.

A crucial point to note now is that the single particle and multiparticle states come in different supermultiplets with different number of fermion zero modes. In our example,  each single particle state is a 16-dimensional short multiplet in four dimensions. This arises from quantizing the four complex fermion zero modes that arise from eight broken supersymmetries or equivalently the zero modes of  Green-Schwarz fermions. In a free theory,  multiparticle state  with $I$ particles, will therefore have $4I$ zero modes. It is a much larger supermultiplet which is an I-fold  tensor product of the 16-dimensional supermultiplet.

Now, to count the states we have to compute an appropriate helicity supertrace in 4d spacetime that soaks up the fermion zero modes. In this example, the correct  helicity supertrace would be $B_4 = \Tr [ (-1)^{2\lambda} (2\lambda)^4]$ where $\lambda$ is the four dimensional helicity. On the worldvolume SCFT this would correspond to a modified elliptic genus $\CE_4$. Because of excessive fermionic zero modes, most of the multiparticle states, even though they surely exist in the spectrum, will not contribute. However, some symmetrized multi-particle states {\it can} contribute because symmetrization will effectively reduce the fermionic zero modes from $4I$ to $4$.

To illustrate this more clearly, consider a simple example of a state with total winding number $3$. This state can be realized either as  a single string  with winding number $3$; or as two strings, one with winding number $1$ and the other with winding number $2$; or as  three strings all with winding number $1$ each.  A  single string  with winding number $3$ will surely contribute because it carries four fermion zero modes. Two strings, one with winding number $1$ and the other with winding number $2$ cannot contribute because they have twice as many zero modes. Three strings all with winding number $1$ each will contribute only when they are sitting on top of each other, all other quantum numbers are identical and they are symmetrized so that the effective number of fermion zero modes is still four.

Adding momentum does not change this basic picture. {\it A priori}, this Hilbert space is a direct sum of  many states of different number of strings with different winding numbers and different momentum distributions such that the total winding and momentum add up to $(w,n)$. However, now  the BPS mass formula prevents certain splittings that were allowed with just winding number. Fermionic zero modes further restrict the allowed states.

For a generic point in  moduli space, the BPS mass formula for a string of momentum and winding $(n,w)$ is a square root of the sum of squares
\be\label{BPSmass}
m = \sqrt{n^2 + C_1 w^2  + C_2 nw}
\ee
where $C_{1,2}$ are real numbers depending on the K\"ahler and complex structure modulus. See for example \cite{Harvey:1995fq}. The mass formula therefore only allows for a BPS splitting $m = \sum m_i $ only if
$(n,w)$ have a common factor, $(n,w)= a(m,v)$. Then the splitting is $n_i = m n'_i, w_i = v w'_i$, with $\sum_i n'_i =  \sum w'_i = a $.
The Hilbert space structure is given by summing over all such common factors. Let us denote the Hilbert space of states with momentum $n$  in a particular long string of length $w$ by $\mathcal{H}(w,n)$. The Hilbert space of BPS states is given by a sum over the common divisors of $n$ and $w$ labeling all the ways in which a string can split:
\bea\label{BPSHilb}
\mathcal{H} & = & \sum_{a|n, a|w} \oplus_{\{s_{r}\}} \otimes_{r>0} \mathrm{Sym}^{s_{r}}\, \mathcal{H}(rv,rm) , \quad \sum_r (r m) s_r = w  \cr
& = & \sum_{a|n, a|w}  \oplus_{\{s_{r} \}}\otimes_{r>0} \mathrm{Sym}^{s_{r}}\, \mathcal{H}(rv,rm), \quad \sum_r r s_r = a.
%& = & \mathcal{H}(\mathrm{Sym}^{w}(X)).
\eea

When there are fermion zero modes in the system, the sum further localizes to those partitions of $a$ which are factorizations as explained in $\S{\ref{theorem}}$ and we get
\bea\label{BPSHilbmore}
\mathcal{H} & = & \sum_{a|n, a|w} \sum_{s|a} \mathrm{Sym}^{s} \, \mathcal{H}(rv,rm) , \quad r s = a , \cr
& = & \sum_{a|n, a|w} \sum_{s|a} \mathrm{Sym}^{s} \, \mathcal{H}(av/s,am/s) , \cr
& = & \sum_{a|n, a|w} \sum_{s|a} \mathrm{Sym}^{s} \, \mathcal{H}(w/s,n/s)  , \cr
& = & \sum_{s|n, s|w} \mathrm{Sym}^{s} \, \mathcal{H}(w/s,n/s) .
\eea

To conclude, this analysis gives us a qualitative understanding of
which KK-momentum bound states can appear in our counting problem.
The essential physics of the sum over $s$, which runs over all divisors of $I = w$, is determined by the fermionic zero modes. The power of $s$ will depend on what modified elliptic genus one is computing. For example, instead of toroidally compactified heterotic string, we could have  considered Type-II or heterotic strings either on $T^4$ or $K3$ with different numbers of worldsheet supersymmetry on the left and right. Depending on the number of fermionic zero modes, one would need to compute appropriate modified elliptic genus $\CE_k$ and accordingly the sum over $s$ will be weighted with $s^{k-1}$. However, in all cases, an s-particle state will contribute only if s divides $I$ and $n$ as in (\ref{modell}). This does not preclude the possibility that other bound states may exist which might have more fermionic zero modes.

\bibliographystyle{utphys}
\bibliography{all}

\end{document}